\documentclass[reqno,12pt,a4]{article}
\usepackage{amsmath, amssymb, slashed,graphicx,color,multicol}
\usepackage{cellspace} 
\usepackage[page,toc]{appendix}
\usepackage[colorlinks=true,urlcolor=black,linkcolor=black,citecolor=black]{hyperref}
\usepackage[utf8]{inputenc}

\newcommand{\ubar} \underline

\newcommand{\Gamme}{\mathsf{\Gamma}}

\newcommand{\F}{\mathrm{F}}

\begin{document}
\title{The gluon condensate in an effective $SU(2)$~Yang--Mills theory}

\author{
A.N. Efremov
\\
CPHT, Ecole Polytechnique, CNRS, Université Paris-Saclay,\\
Route de Saclay, 91128 Palaiseau, France.
\\
\href{mailto:alexander@efremov.fr}{alexander@efremov.fr}
}

\maketitle
\begin{abstract}
We make progress towards a derivation of a low energy effective theory for $SU(2)$~Yang--Mills theory. This low energy action is computed to 1-loop using the renormalization group technique, taking proper care of the Slavnov--Taylor identities in the Maximal Abelian Gauge. After that, we perform the Spin-Charge decomposition in a way proposed by L.D. Faddeev and A.J. Niemi. The resulting action describes a pair of non-linear $O(3)$ and $\sigma$-models interacting with a scalar field. The potential of the scalar field is a Mexican hat and the location of the minima sets the energy scale of solitonic configurations of the $\sigma$-model fields whose excitations correspond to glueball states.
\end{abstract}
\section{Introduction}
The force between nucleons has been studied for many decades. Yang--Mills (YM) theory is a basic building block in the standard model and plausibly can be regarded as a starting point in our quest for an effective low energy model for strong interactions. The Skyrme model is a classic proposal for such an effective theory made by the British physicist T.H.R. Skyrme in his pioneering work “Nonlinear field theory”, published in 1961~\cite{skyrme}. Other models were later proposed by Y.S.~Duan, M.L.~Ge and independently by Y.M.~Cho~\cite{cho80} then later developed further by L.D.~Faddeev, A.J.~Niemi~\cite{fad99} (apparently the idea goes back to a suggestion made by L.D.~Faddeev in 1975). These works are based on a decomposition of the non-abelian gauge potential using the method of abelian projection~\cite{hooft81}, the idea is being that abelian degrees of freedom (corresponding to a $U(1)$ subgroup of the non-abelian gauge group) somehow play an important role at low energies. An alternative so called "Spin-Charge" decomposition is also based on this projection proposed subsequently in~\cite{fad2007,knots2008}.

However a prerequisite for the Spin-Charge decomposition can not be fulfilled in classical YM theory. In the present work, we make modest attempt to bridge this gap. We proceed by integrating out the high energy degrees of freedom in YM theory down to some infrared cutoff $\Lambda$, obtaining the effective action $\Gamma^{\Lambda}$. This calculation is performed with the aid of the renormalization group flow equations~\cite{fhh}, for clarification see the comment below~\eqref{1411g}. Our calculation is performed in the Maximal Abelian Gauge (MAG) including, as usual, gauge fixing and ghost terms into the YM action. However, the end result is gauge invariant in the sense that the effective action for $\Lambda = 0$ would satisfy the Slavnov--Taylor identities~\cite{st2,taylor}. Aiming to obtain an effective model we do not go all the way down to $\Lambda = 0$, because one does not expect a 1-loop calculation, such as ours, to capture the non-perturbative effects that become dominant at low energies. On the other hand the Spin-Charge decomposition becomes singular as the coupling decreases. For these reasons we work with~$\Gamma^\Lambda$ at some small but finite~$\Lambda$. 

This low energy effective action~$\Gamma^\Lambda$ contains, besides the "tree-level" terms, which are already present at the classical level, certain 1-loop correction terms. The most important term is of the form $\delta m^2 A^+ A^-$, where $\delta m^2$ is a mass terms of the order $\Lambda^2$ generated by 1-loop quantum correction, and $A^\pm$ are the
components of the gauge field orthogonal to the abelian $U(1)$-component~$A$. Our finding is that $\delta m^2 < 0$. At this point, we can now perform the Spin-Charge decomposition~\cite{fad2007}, which effectively exchanges $A^\pm$ for a scalar density $\rho$, and two $O(3)$ nonlinear $\sigma$-models described respectively by $\mathbf{p}$, $\mathbf{q}$ and~$\mathbf{n}$ variables. Due to $\delta m^2 < 0$, the stable minimiser of the effective action~$\Gamma^\Lambda$ approaches a non-vanishing value of $\rho=\delta m^2/g^2$ a large spatial distances. This value in turn corresponds to the coupling parameter of the $O(3)$ nonlinear $\sigma$-models.

The key idea is now that the stable configurations of $\mathbf{n}$ are knotted solitons~\cite{fad97,knots2008}. Since the coupling parameter of the corresponding $\sigma$-model sets the scale for the energy of these configurations, the asymptotic value of $\rho$ determines the energy scale for the glueballs.

Calculations with a somewhat similar aim were previously carried out by U.~Ellwanger~\cite{Ellwanger} and H.~Gies~\cite{gies}, with apparently contradicting results. The calculations in~\cite{Ellwanger,gies} were performed in the Lorenz gauge, rather than the MAG as in the present paper. Consequently, it is unclear to us what would be the relevance of these results to the Spin-Charge decomposition which is our main concern.

This paper is organized as follows. We first review the basic setup for the MAG in section~\ref{mag}. After that, we state in section~\ref{sti} the Slavnov--Taylor identities in this gauge. Section~\ref{mass} contains our main result. Here we derive an expression that is helpful to study convexity of the effective action in an efficient way. In section~\ref{model} we explain the relationship of our result with the Spin-Charge decomposition. 

\section{Maximal Abelian Gauge}\label{mag}
We consider SU(2) YM theory with gauge field $\mathsf{A}=\mathsf{A}^i \, \frac{\sigma_i}{2}$ where $\sigma_i$ are the Pauli matrices. For $a \in (+,3,-)$  we  denote by $A^a$ the following components
\begin{align}
A^{\pm}&=\frac{\mathsf{A}^2 \pm i \mathsf{A}^1}{\sqrt{2}}\,,&A=\mathsf{A}^3\,. \label{811a}
\end{align}
The classical YM action can be expanded as
\begin{equation}
L^{YM}=\int d^4 x\sum^4_{n=2}  \mathcal{L}^{YM}_{n},
\end{equation}
where $n$ stands for the number of fields. Integrating by parts we may write
\begin{align}
\mathcal{L}^{YM}_{2}&=\frac{1}{4}F_{\mu \nu} F_{\mu \nu} + \partial_\mu A^+_\nu \partial_\mu A^-_\nu - \partial_\nu A^+_\nu \partial_\mu A^-_\mu ,\\
\mathcal{L}^{YM}_{3}&=- ig A^+_\mu A^-_\nu F_{\mu \nu} + ig A_\mu A^-_\nu F^+_{\mu \nu} - ig A_\mu A^+_\nu F^-_{\mu \nu},\\
\mathcal{L}^{YM}_{4}&=g^2(A_\mu A_\mu A^+_\nu A^-_\nu - A_\mu A^+_\mu A_\nu A^-_\nu) + \frac{g^2}{2}((A^+_\mu A^-_\mu)^2 - A^+_\mu A^+_\mu A^-_\nu A^-_\nu)\,,
\end{align}
where $F^a_{\mu \nu}=\partial_{\mu} A^a_\nu - \partial_\nu A^a_\mu$. 
An infinitesimal gauge transformation has the following form:
\begin{align}
\delta A&=\partial \alpha - ig (A^{+} \alpha^{-} - A^{-} \alpha^{+}),&\delta A^{\pm}&=D^{\pm}\alpha^{\pm} \pm ig\alpha A^{\pm}.\label{811b}
\end{align} 
Here $\alpha^a$ are arbitrary functions, $D^{\pm}$ stands for a covariant derivative 
\begin{equation}
D^{\pm}_\mu A^{\pm}_\nu=\partial_\mu  A^{\pm}_\nu \mp ig A_\mu A^{\pm}_\nu \, 
\end{equation}
with respect to the abelian subgroup~$U(1)$ corresponding to the generator~$\sigma^3$. The field in the MAG is required to be an extremum of $(\mathsf{A}^1)^2 + (\mathsf{A}^2)^2$ over gauge transformations, which coincides with the extremum of $A^+_\nu A^-_\nu$. Substituting~\eqref{811b} into $(\delta A^+_\nu) A^-_\nu + A^+_\nu (\delta A^-_\nu)$ and integrating by parts we obtain
\begin{equation}
0= - \int d^4 x (\alpha^+ D^-_\nu A^-_\nu + \alpha^- D^+_\nu A^+_\nu) \implies D^{\pm}_\nu A^{\pm}_\nu=0.
\end{equation}
The quantity $A^+_\nu A^-_\nu$ is invariant under the residual $U(1)$ gauge transformation 
\begin{equation}
\delta A^{\pm}=\pm ig \alpha A^{\pm}.\label{1911a}
\end{equation}
The abelian symmetry~$U(1)$ can be gauge fixed by imposing the condition $\partial A=0$, i.e. the Lorenz gauge fixing. 
The gauge fixed action is supplemented, as usual, with the Faddeev-Popov term in order to 
account for the Faddeev-Popov determinant, see~\eqref{811d}. The corresponding ghost fields are denoted by
\begin{align}
\mathbf{c}&=(c,c_+,c_-),&\mathbf{\bar{c}}&=(\bar{c},\bar{c}_-,\bar{c}_+),\label{811c}
\end{align}
The Faddeev-Popov  action is
\begin{equation}
L^{FP}=\int d^4 x \sum^4_{n=2} \mathcal{L}^{FP}_n,
\end{equation}
where
\begin{align}
\mathcal{L}^{FP}_2&=\bar{c}_- \partial^2 c_+ + \bar{c}_+ \partial^2 c_- + \bar{c} \partial^2 c\,,\\
\mathcal{L}^{FP}_3&= ig(\partial_\nu \bar{c}_-) A_\nu c_+-ig\bar{c}_- A_\nu (\partial_\nu c_+) - ig (\partial_\nu \bar{c}_+) A_\nu c_- + ig A_\nu \bar{c}_+ (\partial c_-) \nonumber\\
&\quad- ig(\partial_\nu \bar{c})(A^-_\nu c_+ - A^+_\nu c_-) + ig(\bar{c}_- c \partial_\nu A^+_\nu - \bar{c}_+ c \partial_\nu A^-_\nu)\,,\\
\mathcal{L}^{FP}_4&=g^2(\bar{c}_- c A_\nu A^+_\nu + \bar{c}_+ c A_\nu A^-_\nu)+g^2(A^+_\nu A^-_\nu - A_\nu A_\nu) (\bar{c}_- c_+ + \bar{c}_+ c_-) \nonumber\\
&\quad- g^2(A^+_\nu A^+_\nu\bar{c}_- c_- + A^-_\nu A^-_\nu \bar{c}_+ c_+)\,.
\end{align}
The t'Hooft gauge fixing Lagrangian density implementing the MAG and the Lorenz condition is
\begin{equation}
\mathcal{L}^{GF}= \frac{1}{2\xi} (\partial A)^2 + \frac{1}{\xi}(D^+_\nu A^+_\nu)(D^-_\mu A^-_\mu)\,.\label{0502a}
\end{equation}
Actually, the MAG is recovered in the limit $\xi\to 0$. 
Introducing, as usual, auxiliary fields $B^a$, we obtain an equivalent form $\mathcal{L}^{GF}=\mathcal{L}^{GF}_2+\mathcal{L}^{GF}_3$ where
\begin{align}
\mathcal{L}^{GF}_2&=\frac{\xi}{2}B^2 + \xi B^+ B^- - iB^a \partial_\nu A^a_\nu \,,\\
\mathcal{L}^{GF}_3&=gB^{+} A_\nu A^{-}_\nu - gB^{-} A_\nu A^{+}_\nu\,.
\end{align}
Let $\Phi=(\varphi^a, c_a, \bar{c}_a)$, $\varphi^a=(A^a,B^a)$. Denote by $\theta$ a Grassmannian number anti-commuting with $\bar{c}_a$, $c_a$. It is easy to verify that the semi-classical action $L^{YM} + L^{FP} + L^{GF}$ is invariant under the BRST transformation $\delta \Phi=\theta \, s \Phi$:
\begin{align}
s A &= \partial c -ig (A^+ c_+ - A^- c_+),&sc&=ig c_+ c_-,&s \bar{c}_a&=iB^a,\label{1911b}\\
s A^{\pm}&=D^{\pm} c_{\pm} \pm ig c A^{\pm},&sc_{\pm}&=\pm ig c c_\pm,&sB^a&=0.\label{1911c}
\end{align}
Clearly the action is also invariant under the global $U(1)$ symmetry, corresponding to a constant~$\alpha$ in~\eqref{1911a}, and under the Euclidean isometry group. Although our decomposition explicitly breaks the $SU(2)$ invariance there is still a residual discrete symmetry:
\begin{align}
\varphi &\mapsto -\varphi,&c&\mapsto -c,&\bar{c}&\mapsto -\bar{c},\\
\varphi^{\pm} &\mapsto \varphi^{\mp},&c_\pm &\mapsto c_{\mp},&\bar{c}_{\pm}&\mapsto \bar{c}_{\mp}.
\end{align}
The unregularized propagator~$\mathbf{C}$ is obtained by setting $g=0$ in the action:
\begin{equation}
\frac{1}{2}\langle \Phi, \mathbf{C}^{-1} \Phi \rangle=\left(L^{YM} + L^{GF} + L^{FP}\right)\Big|_{g=0}\,.
\end{equation}
The non-vanishing matrix elements of the propagator are:
\begin{align}
\mathbf{C}_{A^a_\mu A^{b}_\nu}&=C_{\mu \nu} \delta_{a \check{b}},&\mathbf{C}_{A^a_\mu B^{b}}&=p_\mu S\delta_{a \check{b}},&\mathbf{C}_{B^a A^{b}_\nu }&=-p_\nu S \delta_{a \check{b}},\\
\mathbf{C}_{c_{a} \bar{c}_b}&=-S\delta_{a \check{b}},&\mathbf{C}_{\bar{c}_{a} c_b}&=S\delta_{a \check{b}}\,,
\end{align}
where the map $b \mapsto \check{b}$ stands for $(+,3,-) \mapsto (-,3,+)$ and 
\begin{align}
C_{\mu \nu}(p)&=\frac{1}{p^2}(\delta_{\mu \nu} + (\xi -1 )\frac{p_\mu p_\nu}{p^2}),&S(p)&=\frac{1}{p^2}\,.
\end{align}
We choose the following regularization:
\begin{align}
\mathbf{C}^{\Lambda \Lambda_0}(p)&=\mathbf{C}(p)\sigma_{\Lambda \Lambda_0}(p^2),\\
\sigma_{\Lambda\Lambda_0}(s^2)&=\sigma_{\Lambda_0}(s^2)-\sigma_{\Lambda}(s^2),&\sigma_{\lambda}(s^2)&=e^{-\left(\frac{s^2}{\lambda^2}\right)^n}\,.
\end{align}
The parameters~$\Lambda$, $\Lambda_0$ are the infrared and ultraviolet cutoffs, $n>0$ is an integer. At large~$n$ the regulator $\sigma_{\lambda}$ approximates a step function.

Introducing the external sources $K=(k^{\check{a}}, \bar{\eta}_{\check{a}}, \eta_{\check{a}})$ where $k^a=(j^a,b^a)$, we define the complex measure $d \mu_{\Lambda \Lambda_0}(\Phi)$ by
\begin{equation} 
e^{-\frac{1}{2}\langle K, \mathbf{C}^{\Lambda \Lambda_0} K \rangle}=\int d \mu_{\Lambda \Lambda_0}(\Phi) e^{i \langle \Phi \cdot K \rangle}.\label{1411e}
\end{equation}
Here $K \cdot \Phi=\Phi \cdot K=\varphi^a k^{\check{a}} + \bar{\eta}_a c_{\check{a}} + \bar{c}_a \eta_{\check{a}}$ is a special notation for the product with anti-ghost on the left. The bosonic part of $\mathbf{C}$ can be diagonalized by a linear map $j^a \mapsto j^{\prime a}=j^a - i \frac{1}{\xi}\partial b$, i.e. $\langle j,\mathbf{C} j \rangle = \langle j^\prime, \mathbf{C}^\prime j^\prime \rangle$, where
\begin{align}
\mathbf{C}^\prime_{A^a_\mu A^{b}_\nu}&=C_{\mu \nu} \delta_{a \check{b}},&\mathbf{C}^\prime_{B^a B^b}&= \frac{1}{\xi} \delta_{a \check{b}},\\
\mathbf{C}^\prime_{c_{a} \bar{c}_b}&=-S\delta_{a \check{b}},&\mathbf{C}^\prime_{\bar{c}_{a} c_b}&=S\delta_{a \check{b}}\,.
\end{align}
Using the diagonalized form we obtain the following decomposition:
\begin{equation}
d \mu_{\Lambda \Lambda_0}(A,B)=d \nu_{\Lambda \Lambda_0} (A) d \upsilon_{\Lambda \Lambda_0}(B - i \frac{1}{\xi} \partial A).
\end{equation}
Here $d \nu_{\Lambda \Lambda_0}$, $d \upsilon_{\Lambda \Lambda_0}$ are Gaussian measures with covariances $C^{\Lambda \Lambda_0}$ and $\frac{1}{\xi} \sigma_{\Lambda \Lambda_0}$. This decomposition allows us to easily obtain the variation of the measure $d \mu$ corresponding to a change of variables $\Phi \mapsto \Phi + \delta \Phi$ using familiar formulas for Gaussian measures. In particular for the infinitesimal variation $\delta \Phi=(\delta A^a, 0, \delta c_a , \delta \bar{c}_a)$ one has
\begin{equation}
\delta[d \mu_{\Lambda \Lambda_0}(\Phi)]=- \langle \delta \Phi, \mathbf{C}^{-1}_{\Lambda \Lambda_0} \Phi \rangle d \mu_{\Lambda \Lambda_0}(\Phi)\,.\label{1311a}
\end{equation}
As a standard procedure we define the partition function
\begin{align} 
Z^{\Lambda \Lambda_0}(K)&=\int d \mu_{\Lambda \Lambda_0}(\Phi) e^{-L^{\Lambda_0 \Lambda_0} + \langle K \cdot \Phi \rangle}\,,\label{1311b}\\
L^{\Lambda_0 \Lambda_0}&=\sum \limits_{n \in \{3,4\}} \left(L^{YM}_n + L^{GF}_n + L^{FP}_n \right) + L^{\Lambda_0}_{ct}\,,
\end{align}
where $L^{\Lambda_0}_{ct}$ includes all counterterms. The generating functional for Connected Schwinger functions is defined by~$W=\log Z$. Performing the Legendre transform we obtain the effective action 
\begin{equation}
\Gamma^{\Lambda \Lambda_0}(\Phi)=\langle \Phi \cdot K \rangle  -  W^{\Lambda \Lambda_0}(K)\,.\label{1411d}
\end{equation}
We close this list of definitions by auxiliary functionals $\Gamme^{\Lambda \Lambda_0}$, $\F^{\Lambda \Lambda_0}$:
\begin{align}
\Gamme^{\Lambda \Lambda_0}&=\Gamma^{\Lambda \Lambda_0} - \frac{1}{2}\langle \Phi \mathbf{C}^{-1}_{\Lambda \Lambda_0} \Phi \rangle ,&\F^{\Lambda \Lambda_0}&=\frac{1}{2}\langle \Phi \mathbf{C}^{-1}_{0 \Lambda_0} \Phi \rangle + \Gamme^{\Lambda \Lambda_0}.\label{1411a}
\end{align}

\section{Slavnov--Taylor identities}\label{sti}
In this section we derive the Slavnov--Taylor identities in the presence of the infrared cutoff~$\Lambda$. These identities hold for a finite $\Lambda$.
As usual, the introduction of these identities encodes the local gauge invariance of the theory and effectively constrains the appropriate choice of renormalization conditions. 
Details can be found in~\cite{egk}.

First we add to the Lagrangian the renormalized BRST insertions $\psi^a$, $\Omega^a$ (see~\eqref{1911b}, \eqref{1911c}) which are invariant under global $U(1)$ and discrete symmetries
\begin{equation}
L^{\Lambda_0 \Lambda_0}_{vst}=L^{\Lambda_0 \Lambda_0}+\langle \gamma^a,\psi^{\check{a}} \rangle + \langle \omega^a,\Omega^{\check{a}} \rangle,
\end{equation}
where $\gamma^a$, $\omega^a$ are auxiliary sources having ghost number $-1$ and $-2$,
\begin{align}
\psi&= R^{\Lambda_0}_1 \partial c + ig R^{\Lambda_0}_2(A^+ c_- - A^- c_+),&\Omega&=ig R_3 c_- c_+\,,\\
\psi^{\pm}&=R^{\Lambda_0}_4 \partial c_{\pm} \mp ig R^{\Lambda_0}_5 A c_{\pm} \pm igR^{\Lambda_0}_6 cA^{\pm},&\Omega^\pm&=\pm ig R_7 c_\pm c\,.\label{0602a}
\end{align}
Here $R^{\Lambda_0}_{i}=1+ r^{\Lambda_0}_i$ where $r^{\Lambda_0}_i$ are fixed by renormalization conditions. Performing in \eqref{1311b} the change of variables $\Phi \to \Phi + \delta \Phi$ where 
\begin{align}
\delta A^a&=\theta \sigma_{0 \Lambda_0}\psi^a,&\delta B^a&=0,&\delta c_a&=-\theta \sigma_{0 \Lambda_0} \Omega^a,&\delta \bar{c}_a=\theta i \sigma_{0 \Lambda_0} B^a,
\end{align}
and using \eqref{1311a} we obtain the identity
\begin{multline}
\int d \mu_{\Lambda \Lambda_0} e^{-L^{\Lambda_0 \Lambda_0} + \langle \Phi \cdot K\rangle } \Big\{ \Upsilon^{\Lambda_0} + \langle \begin{pmatrix}\psi^a,0,-\Omega^a,iB^a\end{pmatrix}, \delta \mathbf{C}^{-1}_{ab \, \Lambda \Lambda_0} \Phi^b\rangle \\= \langle K^{\check{a}},\sigma_{0 \Lambda_0}\begin{pmatrix}\psi^a, 0, \Omega^a, iB^a \end{pmatrix} \rangle \Big\},\label{1311c}
\end{multline}
where $\delta \mathbf{C}^{-1}_{\Lambda \Lambda_0}= \mathbf{C}^{-1} \delta \sigma_{\Lambda \Lambda_0}$, $\delta \sigma_{\Lambda \Lambda_0}=\sigma_{\Lambda} \sigma^{-1}_{\Lambda \Lambda_0}$ and
\begin{align}
\Upsilon^{\Lambda_0}&= \langle \psi^a, \sigma_{0 \Lambda_0}\frac{ \delta L^{\Lambda_0 \Lambda_0}}{\delta A^a}\rangle + i \langle B^a, \sigma_{0 \Lambda_0}\frac{ \delta L^{\Lambda_0 \Lambda_0}}{\delta \bar{c}_a}\rangle - \langle \Omega^a, \sigma_{0 \Lambda_0}\frac{ \delta L^{\Lambda_0 \Lambda_0}}{\delta c_a}\rangle \nonumber\\
&\quad+\langle \begin{pmatrix}\psi^a,0,-\Omega^a,iB^a \end{pmatrix} \mathbf{C}^{-1}_{ab} \Phi^b \rangle\,.
\end{align}
Let $L^{\Lambda_0 \Lambda_0}_{aux}=L^{\Lambda_0 \Lambda_0}_{vst} + \rho \Upsilon$ where $\rho$ is an auxiliary source with ghost number~$-1$. Since the functionals $L^{\Lambda_0  \Lambda_0}_{aux}$ and $L^{\Lambda_0  \Lambda_0}_{vst}$ at $(\gamma^a,\omega^a,\rho)=0$ coincide with the original~$L^{\Lambda_0 \Lambda_0}$ we can substitute the composite insertions and $\Phi$ in~\eqref{1311c} with derivatives with respect to their sources. This gives an identity for $W^{\Lambda \Lambda_0}_{aux}$ at vanishing~$\rho$,~$\omega$,~$\gamma$:
\begin{multline}
W_1 + \langle \begin{pmatrix}W_{\gamma^a},0,W_{\omega^a},-i\dfrac{\delta W}{\delta b^a} \end{pmatrix} \delta \mathbf{C}^{-1}_{\check{a} \check{b} \, \Lambda \Lambda_0} \begin{pmatrix}\dfrac{\delta W}{\delta j^b},\dfrac{\delta W}{\delta b^b},\dfrac{\delta W}{\delta \bar{\eta}_b},\dfrac{\delta W}{\delta \eta_b}\end{pmatrix} \rangle\\=\langle K^a, \sigma_{\Lambda \Lambda_0}\begin{pmatrix} W_{\gamma^a}, 0, W_{\omega^a}, -i\dfrac{\delta W}{\delta b^a}  \end{pmatrix} \rangle - \Delta^{\Lambda \Lambda_0},\label{1411b}
\end{multline}
where 
\begin{align}
W^{\Lambda \Lambda_0}_1 &= \frac{\partial W^{\Lambda \Lambda_0}_{aux}}{\partial \rho},&W^{\Lambda \Lambda_0}_\gamma&= \frac{\delta W^{\Lambda \Lambda_0}_{aux}}{\delta \gamma},&W_\omega&= \frac{\delta W^{\Lambda \Lambda_0}_{aux}}{\delta \omega}\,,
\end{align}
\begin{equation}
\Delta^{\Lambda \Lambda_0}= \langle \begin{pmatrix}\dfrac{\delta}{\delta \gamma^a},0,\dfrac{\delta}{\delta \omega^a},- i\dfrac{\delta}{\delta b^a}\end{pmatrix} \delta \mathbf{C}^{-1}_{\check{a} \check{b} \, \Lambda \Lambda_0} \begin{pmatrix}\dfrac{\delta }{\delta j^b},\dfrac{\delta }{\delta b^b},\dfrac{\delta }{\delta \bar{\eta}_b},\dfrac{\delta }{\delta \eta_b}\end{pmatrix} \rangle W^{\Lambda \Lambda_0}\,.\label{1511a}
\end{equation}
Performing in~\eqref{1411b} the Legendre transform~\eqref{1411d} and using definition~\eqref{1411a} we obtain the Slavnov--Taylor identities at a non-vanishing value of~$\Lambda$
\begin{equation}
\Gamma^{\Lambda \Lambda_0}_1=\langle \frac{\delta \F^{\Lambda \Lambda_0}}{ \delta A^a} \sigma_{0 \Lambda_0} \Gamma^{\Lambda \Lambda_0}_{\gamma^{\check{a}}} \rangle -\langle\frac{\delta \F^{\Lambda \Lambda_0}}{\delta c_a} \sigma_{0 \Lambda_0} \Gamma^{\Lambda \Lambda_0}_{\omega^{\check{a}}}\rangle + i\langle B^a \sigma_{0 \Lambda_0} \frac{\delta \F^{\Lambda \Lambda_0}}{\delta \bar{c}_a} \rangle + \Delta^{\Lambda \Lambda_0}\;.\label{1411g}
\end{equation}
Without giving a proof we state that $\lim \limits_{\Lambda_0 \to \infty} \Gamma^{\Lambda \Lambda_0}_1=0$. Detailed bounds on this term establishing this claim
can be found in recent works~\cite{egk,fhh}.

An important simplification comes from linearity of the Lorenz gauge fixing condition for the $U(1)$-component~$A$. It is easy to see that
\begin{align}
W^{\Lambda \Lambda_0}&= \frac{1}{2} \langle b, \mathbf{C}^{\Lambda \Lambda_0}_{BB} b \rangle + \tilde{W}^{\Lambda \Lambda_0}\Big(j - \frac{\partial b}{\xi}\Big)\,,&\mathbf{C}^{\Lambda \Lambda_0}_{BB}&=\frac{1}{\xi}\sigma_{\Lambda \Lambda_0}\,.\label{1411f}
\end{align}
Here to define the measure~$d\tilde{\mu}(\tilde{\Phi})$ corresponding to $\tilde{W}(\tilde{K})$ we put~$b=0$ in~\eqref{1411e}. In other words the measure~$d\tilde{\mu}(\tilde{\Phi})$ is obtained from~$d \mu(\Phi)$ by integrating out~$B$. Performing in~\eqref{1411f} the Legendre transform we get
\begin{equation}
\Gamma^{\Lambda \Lambda_0}(\Phi)=\frac{1}{2} \langle (B-i\frac{\partial A}{\xi})\mathbf{C}^{-1}_{BB \, \Lambda \Lambda_0}(B-i\frac{\partial A}{\xi})\rangle + \tilde{\Gamma}^{\Lambda \Lambda_0}(\tilde{\Phi})\,.
\end{equation}
Substituting into definition~\eqref{1411a} above equation we obtain $\Gamme^{\Lambda \Lambda_0}(\Phi)=\tilde{\Gamme}^{\Lambda \Lambda_0}(\tilde{\Phi})$ and thus
\begin{equation}
\F^{\Lambda \Lambda_0}(\Phi)=\frac{1}{2}\langle \Phi \mathbf{C}^{-1}_{0 \Lambda_0} \Phi \rangle+ \tilde{\Gamme}^{\Lambda \Lambda_0}(\tilde{\Phi})\,.\label{1411h}
\end{equation}
It follows that the dependence on the variable $B$ is rather trivial and captured by the first term in~\eqref{1411h}. For this reason in the following we consider only the tilde effective action $\tilde{\Gamme}^{\Lambda \Lambda_0}(\tilde{\Phi})$ corresponding to the measure $d\tilde{\mu}(\tilde{\Phi})$, i.e. without the field~$B$ and source~$b$. To shorten our notation, we there will omit from now on again the tilde symbol, with the understanding that all quantities 
are actually given for the reduced set of the variables, i.e. $A,$ $A^\pm$, $B^\pm$, $c$, $\bar{c}$, $c_\pm$, $\bar{c}_\pm$. 

After substituting~\eqref{1411h} into~\eqref{1411g} the violated Slavnov--Taylor identities take the form
\begin{align}
\lim \limits_{\Lambda_0 \to \infty}& \Big(\sigma_{0 \Lambda_0} \frac{\delta \F}{\delta \bar{c}} - \partial \Gamma_\gamma\Big) = 0,\\
\lim \limits_{\Lambda_0 \to \infty}& \Big(\langle \frac{\delta \F}{\delta A^a} \sigma_{0 \Lambda_0} \Gamma_{\gamma^{\check{a}}}\rangle  - \langle \frac{\delta \F}{\delta c_a} \sigma_{0 \Lambda_0} \Gamma_{\omega^{\check{a}}}\rangle\nonumber\\ &- \frac{1}{\xi}\langle \partial A \sigma_{0 \Lambda_0} \frac{\delta \F}{\delta \bar{c}}\rangle + i\langle B^\pm \sigma_{0 \Lambda_0} \frac{\delta \F}{\delta \bar{c}_\pm}\rangle + \Delta^{\Lambda \Lambda_0}\Big) =  0,\label{1511d}\\
\Delta^{\Lambda \Lambda_0}=&-\langle \sigma_\Lambda (1 + \Gamme_2 \hat{\mathbf{C}})^{-1}_{A^a \Phi} \frac{\delta \Gamma_{\gamma^{\check{a}}}}{\delta \Phi} \rangle - \langle \sigma_\Lambda \partial (1 + \Gamme_2 \hat{\mathbf{C}})^{-1}_{A^a c_a} \rangle\nonumber\\
&+\langle \sigma_\Lambda (1 + \Gamme_2 \hat{\mathbf{C}})^{-1}_{c_a \Phi} \frac{\delta \Gamma_{\omega^{\check{a}}}}{\delta \Phi} \rangle,\label{1511c}\\
\Gamme_2=&\frac{\delta^2 \Gamme}{\delta \Phi \delta \Phi^\prime },\quad \hat{\mathbf{C}}=\hat{\mathbf{1}}\mathbf{C},\quad \hat{\mathbf{1}}=\left\{\begin{matrix}\delta_{\varphi^a \varphi^b}\\-\delta_{c_a c_b}\\-\delta_{\bar{c}_a \bar{c}_b}\end{matrix}\right.\label{0602e}
\end{align}

\section{Convexity of the effective action}\label{mass}

In this section, we use the results of the previous section, specifically equations \eqref{1511d}-\eqref{0602e} in order to obtain the effective action in the MAG (satisfying the Slavnov--Taylor identities) in the 1-loop approximation. Here, by the loop expansion we mean, as usual, an expansion in $\hbar$, i.e. 
$\Gamme=\sum_l \hbar^l \Gamme_l$.

The effective action~$\Gamme^{\Lambda \Lambda_0}$ can be written as
\begin{equation}
\Gamme^{\Lambda \Lambda_0}=\langle A^+_\mu (p^2\delta_{\mu \nu } - p_\mu p_\nu)\Sigma^{\Lambda \Lambda_0} A^-_\nu \rangle  +  \langle  A^+ \delta m^{2}_{\Lambda \Lambda_0} A^- \rangle+...\label{1511b}
\end{equation}
where we only keep explicitly the terms quadratic terms in $A^{\pm}$. (Note again that for shortness of notations we drop the tilde.)
We focus specifically on the last "mass-term" because it appears only at 1-loop and therefore induces a qualitative difference to the 0-loop order (tree level $=$ classical)
action. Indeed, according to definition of~$\Gamme^{\Lambda \Lambda_0}$ in~\eqref{1411a} both terms in~\eqref{1511b} vanish at tree level,~$l=0$: The first one since the effective action by our conventions does not 
include the quadratic kinetic terms~\eqref{1411a}, and the second one because the classical action has no mass terms for the gauge fields. 

To derive the value of $\delta m^2$, we consider the vertex function $\Gamma^{A^+ c_-}_1$. From identity~\eqref{1411g} we have 
\begin{equation}
\Gamma^{A^+ c_-}_1=\F^{A^+ A^-} \sigma_{0 \Lambda_0} \Gamma^{c_-}_{\gamma^{+}} + \Delta^{A^+_\beta c_-}.
\end{equation}
Using the violated Slavnov--Taylor identities, see~\eqref{1511d}, and then taking the partial derivative at the point $p=0$ we obtain
\begin{equation}
\F^{A^+ A^-} \partial_p \Gamma^{c_-}_{\gamma^{+}} + \partial_p \Delta^{A^+_\beta c_-}(p) \Big|_{p=0} =0.
\end{equation}
Finally, using~\eqref{1411h},~\eqref{1511b} and~\eqref{0602a} we get the identity
\begin{equation}
\delta_{\beta \alpha}  \delta m^2  R_4  = -i \frac{\partial \Delta^{A^+_\beta c_-}(p)}{\partial p_\alpha}\Big|_{p=0}.
\end{equation}
Performing the loop expansion on the left hand side we have 
\begin{equation}
\delta m^2_{l=0} R_{4;l=1} + \delta m^2_{l=1} R_{4;l=0}=\delta m^2_{l=1}.
\end{equation}
To compute the right hand side at 1-loop we observe that $\Delta_{l=1}$ involves~$W_{l=0}$. It can be easily checked by restoring the Planck's constant in~\eqref{1511a}:
\begin{align}
K &\mapsto \frac{1}{\hbar}K,& \gamma &\mapsto \frac{1}{\hbar} \gamma,&\omega &\mapsto \frac{1}{\hbar} \omega,&W &\mapsto \frac{1}{\hbar}W.
\end{align}
Thus, substituting the tree level for~$\Gamme$ in~\eqref{1511c} we obtain
\begin{align}
\Delta^{A^+_\beta c_-}&=i g^2 \int \frac{d^4 \ell}{(2\pi)^4} \sigma_{\Lambda}(\ell^2)\left( (9\ell_\beta + (2\xi -1)p_\beta) S^{\Lambda \infty}_{\ell-p} -2p_\alpha C^{\Lambda \infty}_{\alpha \beta, \ell-p} \right).
\end{align}
This equation gives
\begin{equation}
\delta m^2=g^2 \frac{\Lambda^2 \pi^2}{2(2 \pi)^4}\Gamma(1 + \frac{1}{n})\left((3 \xi +4)(1 - \frac{1}{2^{\frac{1}{n}}}) - \frac{9}{2^{1+\frac{1}{n}} } \right)\label{1911d}. 
\end{equation}
This is the main result of our perturbative analysis. 

There are two independent limits for~\eqref{1911d}. The first corresponds to the MAG 
\begin{align}
\xi&\to 0&\delta m^2&=- g^2 \frac{\Lambda^2 \pi^2}{2(2 \pi)^4}\Gamma(1 + \frac{1}{n})\frac{1 + 8(2- 2^{\frac{1}{n}})}{2^{1+\frac{1}{n}} }\,,
\end{align}
and the second corresponds to a sharp cutoff in the momentum space, i.e. it describes an effective model,
\begin{align}
n&\to \infty&\delta m^2&=- g^2 \frac{\Lambda^2 \pi^2}{2(2 \pi)^4}\frac{9}{2}\,.\label{402a}
\end{align}
In both cases $\delta m^2<0$.

\section{Effective $\sigma$-model}\label{model}
We would like to illustrate the main implication of the negative sign of the mass term obtained in~\eqref{402a}. First we need to briefly introduce the Spin-Charge decomposition proposed by L.D. Faddeev and A.J. Niemi~\cite{fad2007}. Given an orthonormal basis $\mathsf{e}_i$ in the plane $\mathsf{span}(\mathsf{A}^1_\nu, \mathsf{A}^2_\nu)$ the authors define two complex functions~$\psi_i$, a density~$\rho$ and a vector $\vec{t}$
\begin{align}
A^+_\nu&=\psi_1 e_\nu + \psi_2 \bar{e}_\nu,&e &=\mathsf{e}_1 + i\mathsf{e}_2,&\mathsf{e}_i \mathsf{e}_j&=\delta_{ij},\\
A^-_\nu&=\psi^*_2 e_\nu + \psi^*_1 \bar{e}_\nu,&\rho^2&=|\psi_1|^2 + |\psi_2|^2,&\vec{t}&=\frac{1}{\rho^2}\begin{pmatrix}\psi_1^*,\psi^*_2 \end{pmatrix}\vec{\sigma}\begin{pmatrix}\psi_1\\\psi_2 \end{pmatrix}\,.
\end{align}
Here $\psi_1$, $\psi_2$ transform as $\psi_i \to e^{i g\alpha}\psi_i$ under the action of the $U_c(1) \equiv U(1)$ group. In this decomposition the complex fields~$\psi_i$ are equally charged with respect to the $U_c(1)$ group but the complex vector field~$e_\nu$ and vector~$\vec{t}$ are neutral. In this sense the decomposition describes a separation between the charge and the spin of the components $A^{\pm}_\nu$. Furthermore the decomposition is invariant under an internal~$U_i(1)$ symmetry:
\begin{align}
e &\to e^{-i \lambda} e,&\psi_1 &\to e^{i \lambda} \psi_1,&\psi_2 &\to e^{-i \lambda} \psi_2,&t_\pm & \to e^{\mp 2i \lambda} t_\pm,
\end{align}
where $t_\pm=t_1 \pm i t_2$. It is convenient to write the effective action in $U_i(1) \times U_c(1)$ invariant variables.
\begin{align}
H_{\mu \nu}&=\frac{i}{2}(e_\mu \bar{e}_\nu - e_\nu \bar{e}_\mu),&\mathbf{p}_i&=H_{0 i},&\mathbf{q}_i&=\frac{1}{2}\epsilon_{ijk}H_{jk}\\
\hat{e}&=\begin{pmatrix}\frac{2\mathbf{p} \times \mathbf{q} +i \mathbf{p}  }{\sqrt{2} |\mathbf{p}|}, \sqrt{2} |\mathbf{p}|\end{pmatrix},&\hat{C}_\mu &=i \hat{\bar{e}} \partial_\mu \hat{e}.
\end{align}
Here $e=e^{i \eta} \hat{e}$ where $\eta$ is a phase shift. The variables~$\mathbf{p}$, $\mathbf{q}$ are subject to the constraint~$\mathbf{p}^2 + \mathbf{q}^2=\frac{1}{4}$. Moreover instead of the vector~$\vec{t}$ the authors define a new vector $\mathbf{n}_{\pm}=e^{2 i \eta} t_{\pm}$, $\mathbf{n}_3=t_3$. Clearly $\mathbf{p}$, $\mathbf{q}$, $\mathbf{n}$, $\rho$ are $U_i(1)\times U_c(1)$ invariant variables. Defining $g^2 \mu^2_\Lambda =- \delta m^2$ we have
\begin{align}
(A^+_\mu A^-_\mu)^2 - A^+_\mu A^+_\mu A^-_\nu A^-_\nu&=(\rho^2 \mathbf{n}_3)^2,&\delta m^2 A^+A^-&=-g^2 \mu^2_\Lambda \rho^2\,.
\end{align}
Since the Lorenz gauge fixing, $\partial A$ in~\eqref{0502a}, breaks explicitly the $U_c$ invariance we single out this condition into a separate term,
\begin{equation}
\mathcal{L}^{GF}=\frac{1}{2 \xi} (\partial A)^2 + \mathcal{L}^{MAG}.
\end{equation}
Choosing the Feynman gauge, i.e. $\xi=1$, and using~\eqref{402a} we have $U_c \times U_i$ invariant density 
\begin{align}
\mathcal{L}^{YM + MAG}&=(\partial \rho)^2 - g^2 \mu^2_\Lambda \rho^2 + \frac{g^2}{2} (\mathbf{n}_3\rho^2)^2 + \frac{\rho^2}{2}((\partial \mathbf{p})^2 + (\partial \mathbf{q})^2) + \frac{\rho^2}{4}(D^C \mathbf{n})^2 \nonumber\\
&\quad  + \frac{1}{4}(F_{\mu\nu})^2  + g^2 \rho^2 J^2 + \frac{\rho^2}{2}(\mathbf{n}_+ (\partial_\mu \hat{e}^*_\nu)^2 + \mathbf{n}_- (\partial_\mu \hat{e}_\nu)^2).
\end{align}
Here $J_\nu$ is a $U_i(1)\times U_c(1)$ invariant vector which can be found in~\cite{fad2007}. 

One can now ask for classical minimisers of the corresponding  functional. It is reasonable to impose that at large spatial distances a solution for the unit vector~$\mathbf{n}$ is $SO(4)$ invariant. However the components $\mathbf{n}_\pm$~are not invariant. It implies that its  asymptotes are $\mathbf{n}_3=\pm 1$. Then 
at large distance, the potential for the $\rho$-field is of Mexican hat type, 
\begin{equation}
(\partial \rho)^2 - g^2 \mu^2_\Lambda \rho^2 + \frac{g^2}{2} \rho^4\,. \label{2105a}
\end{equation}
Consequently, at large distance, $\rho$ has to go to a non-zero constant. This constant appears then as a coupling for the nonlinear $\sigma$-models for the variables ${\bf p,q}$ and ${\bf n}$, respectively. A simpler version of such $\sigma$-models was investigated numerically~\cite{fad97,sutcliffe2007}. These investigations suggest the existence of knotted i.e. topologically nontrivial solutions that locally minimize the action. If these were to exist also for the above action, they would arguably correspond to non-perturbative excitations of the quantum YM vacuum state. 

One could attempt to interpret the result in a way similar to previous works~\cite{zakh,kondo,kondo2}. However in these works the mass term is calculated at $\Lambda=0$ and corresponds at a qualitative level rather to a non-zero constant for the scalar density~$\rho$ which minimizes the action~\eqref{2105a} than to the mass we have obtained in~\eqref{402a}.  Since it is known that the coupling constant of $O(3)$ nonlinear $\sigma$-model decrease as the spatial distance increases, see, e.g.,~\cite{codello} one can view the decomposition as a leading order approximation.

\section{Acknowledgments}
I want to thank Prof. Ch. Kopper, Ecole Polytechnique, France and Dr. R. Guida, IPhT, CEA, France for reading the manuscript, useful remarks, and numerous discussions.

The research was supported in part by a two month postdoctoral grant from Ecole Polytechnique, France. 

I thank the Institute for Theoretical Physics at the University of Leipzig, Germany for the financial support during my work on the manuscript.

I express my gratitude to Prof. S. Hollands, the University of Leipzig, Germany for warm hospitality, valuable remarks and help.  

\appendix
\section{Faddeev--Popov determinant}
We consider a variation of the gauge fixing conditions under the action of gauge transformation~\eqref{811b}
\begin{equation}
\delta \begin{pmatrix}D^+_\nu A^+_\nu\\D^-_\nu A^-_\nu\\\partial_\nu A_\nu\end{pmatrix}=M \begin{pmatrix}\alpha^+\\\alpha^-\\\alpha\end{pmatrix},
\end{equation}
where
\begin{equation}
M=\begin{pmatrix} D^+_\nu D^+_\nu + g^2 A^+_\nu A^-_\nu&-g^2 A^+_\nu A^+_\nu&igD^+_\nu A^+_\nu\\
-g^2 A^-_\nu A^-_\nu& D^-_\nu D^-_\nu + g^2 A^-_\nu A^+_\nu&-ig D^-_\nu A^-_\nu\\
ig \partial_\nu A^-_\nu& -ig \partial_\nu A^+_\nu& \partial^2\end{pmatrix}
\end{equation}
We define the determinat of $-M$ as follows:
\begin{equation}
\det( -M)= \lim \limits_{N \to \infty} \int \prod \limits^N_{i=1} d \mathbf{c}_i d\mathbf{\bar{c}}_i \;e^{-\mathbf{\bar{c}}_i M_{ij} \mathbf{c}_j}.\label{811d}
\end{equation}
Here $M_{ij}$, $\mathbf{c}_i$, $\mathbf{\bar{c}}_i$ are the corresponding components of $M$, $\mathbf{c}$, $\mathbf{\bar{c}}$ in a finite dimensional basis $\{f^a_i\}^{N}_{i=1}$.

\bibliographystyle{unsrt}
\bibliography{mag}
\end{document}